\documentclass[journal]{IEEEtran}
\usepackage{amssymb}
\usepackage{amsbsy}
\usepackage{amsmath}
\usepackage{amsthm} 
\usepackage{setspace}
\usepackage{epsfig}
\usepackage{cite}
\usepackage{graphics}
\usepackage{epstopdf}
\usepackage{mathrsfs}
\usepackage[above]{placeins}
\usepackage{algorithm}
\usepackage{algpseudocode}
\usepackage{multirow}
\usepackage{algorithmicx}

\begin{document}
%
\title{Decoding by Sampling --- Part II: \\
Derandomization and Soft-Output Decoding}
%
%
%

\author{Zheng~Wang,
        Shuiyin~Liu,~\IEEEmembership{Member, IEEE}
        and~Cong~Ling,~\IEEEmembership{Member, IEEE}
\thanks{This work was presented in part at the IEEE Information Theory Workshop 2012, Lausanne, Switzerland,
September 2012.

Z. Wang and C. Ling are with Department of Electrical and Electronic Engineering, Imperial
College London, London SW7 2AZ, United Kingdom (e-mail: z.wang10@imperial.ac.uk, cling@ieee.org).

S. Liu is with the Department of Electrical and Computer Systems Engineering, Monash University, Melbourne VIC 3180, Australia (e-mail: shuiyin.liu@monash.edu).}}

\maketitle

\begin{abstract}

In this paper, a derandomized algorithm for sampling decoding is proposed
to achieve near-optimal performance in lattice decoding. By setting a
probability threshold to sample candidates, the whole sampling procedure
becomes deterministic, which brings considerable performance improvement
and complexity reduction over to the randomized sampling. Moreover, the
upper bound on the sample size $K$, which corresponds
to near-maximum likelihood (ML) performance, is derived. We also find that
the proposed algorithm can be used as
an efficient tool to implement soft-output decoding in multiple-input
multiple-output (MIMO) systems. An upper bound of the sphere radius $R$ in
list sphere decoding (LSD) is derived.
Based on it, we demonstrate that the derandomized sampling algorithm is capable of
achieving near-maximum \emph{a posteriori} (MAP) performance.
Simulation results show that near-optimum performance can be achieved
by a moderate size $K$ in both lattice decoding and soft-output
decoding.

\end{abstract}


%
\IEEEpeerreviewmaketitle

\section{Introduction}

\IEEEPARstart{A}{s} one of the core problems of lattices, the closest vector problem
(CVP) has wide applications in number theory, cryptography, and
communications. In \cite{Babai}, the lattice reduction technique was introduced to
solve CVP approximately. Its key idea is replacing the original
lattice by an equivalent one with a shorter basis, which greatly improves the performance of
suboptimal decoding schemes like successive interference cancelation (SIC).
Since then, a number of improved decoding schemes
based on the lattice reduction have been proposed
\cite{Luzzi,WubbenMMSE,Agrell2002,Gan2009}. In multiple-input
multiple-output (MIMO) communications,
it has been shown in
\cite{JaldenDMTJ} that minimum mean-square error (MMSE) decoding
based on the lattice reduction achieves the optimal diversity and
multiplexing trade-off. However, the performance gap between maximum-likelihood
(ML) decoding and lattice-reduction-aided decoding is
still substantial especially in high-dimensional systems \cite{CongProxity,DamenDetectionSearch}.

On the other hand, in order to achieve near-capacity performance over MIMO
channels, bit-interleaved coded modulation (BICM) and iterative detection
and decoding (IDD) are well accepted, where the extrinsic information
calculated by \emph{a} \emph{priori} probability (APP) detector is taken
into account to produce the soft decisions \cite{Hochwald}.
As the key ingredient of IDD receivers, the calculation of APP is usually performed
by a log-likelihood ratio (LLR) value via maximum \emph{a posteriori} (MAP)
algorithm, whose complexity increases exponentially with the
number of transmit antennas and the constellation size.
In \cite{Hochwald}, a modified sphere decoding (SD)
algorithm referred to as list sphere decoding (LSD) was given. By resorting to
a list of lattice points within a certain sphere radius, it achieves
an approximation of the MAP performance while maintaining affordable complexity.
However, the exponentially increased complexity
is always a big problem in LSD especially for high-dimensional systems.
Based on LSD, a number of approaches resorting to lattice reduction
were proposed to further reduce the complexity burden or improve
the performance \cite{Hassibisoft,Silvolasoft,MaxiaoliSoft,Millinersoft,BoutrosSoft}.
Unfortunately, none of them give the explicit size of the sphere radius
when the decoder approaches near-MAP performance, making it still an open question.
In \cite{Leesoft}, a LSD-based probabilistic tree pruning algorithm was proposed
with a lower bound constraint of the sphere radius. However, to
fix that initial sphere radius, sphere decoding is still required
as a preprocessing stage making it impractical in high dimensions.

Recently, randomized sampling decoding has been proposed in \cite{CongRandom}
to narrow the gap between lattice-reduction-aided decoding and sphere
decoding. As a randomized version of SIC, it
applies Klein's sampling technique \cite{Klein} to randomly sample
lattice points from a Gaussian-like distribution  and chooses the
closest one among all the samples. However, because of randomization,
there are two inherent issues in random sampling. One is inevitable
repetitions in the sampling process leading to
unnecessary complexity, while the other one is inevitable performance loss
since some lattice points can be missed during the sampling. Although Klein
mentioned a derandomized algorithm very briefly in \cite{Klein}, it does not seem to allow
for an efficient implementation. In \cite{CongRandom}, the randomized
sampling algorithm was also extended to soft-output decoding in MIMO
systems. Although it could achieve remarkable performance gain with
polynomial complexity, it still suffers from these two issues.

In this paper, to overcome these two problems caused by randomization,
we propose a new kind of sampling algorithm referred to as
derandomized sampling decoding. With a sample size $K$ set
initially, candidate points are sampled deterministically according
to a threshold we define. As randomization is removed, derandomized
sampling decoding shows great potential in both performance
and complexity. To further exploit it, its optimum decoding radius, which
is defined in bounded distance decoding (BDD) as a sphere radius that the
lattice point within this radius will be decoded correctly, is derived. Furthermore,
the upper bound on $K$ with respect to near-ML performance is given, by
varying $K$, the decoder enjoys a flexible trade-off between performance
and complexity in lattice decoding.

We then extend derandomized sampling algorithm to soft-output decoding
in MIMO systems. Since the randomization during samplings is removed, it operates
as an approximation scheme like LSD but generates the candidate list by sampling,
which is more efficient and easier to implement. Although samplings are performed
over the entire lattice, lattice points with large sampling probabilities are quite
likely to be sampled, which means the final candidate list tends to be comprised
of a number of lattice points around the closest lattice point. The upper bound
of the sphere radius $R$ in LSD is also derived. Then based on the proposed derandomized
sampling algorithm, the trade-off between performance and complexity in soft-output
decoding is established by adjusting the sample size $K$.

The rest of this paper is organized as follows. Section II presents
the system model and briefly reviews the randomized sampling algorithm in lattice decoding.
In Section III, derandomized sampling algorithm is proposed, followed by performance
analysis and optimization. In Section IV, the proposed algorithm is extended
to soft-output decoding. Simulation results are presented and evaluated in Section V.
Finally, Section VI concludes the paper.

\emph{Notation:} Matrices and column vectors are denoted by upper
and lowercase boldface letters, and the transpose, inverse, pseudoinverse
of a matrix $\mathbf{B}$ by $\mathbf{B}^T, \mathbf{B}^{-1},$ and
$\mathbf{B}^{\dag}$, respectively. We use $\mathbf{b}_i$ for the $i$th
column of the matrix $\mathbf{B}$, $b_{i,j}$ for the entry in the $i$th row
and $j$th column of the matrix $\mathbf{B}$. $\lceil x \rfloor$ denotes rounding to
the integer closest to $x$. If $x$ is a complex number, $\lceil x \rfloor$
rounds the real and imaginary parts separately. Finally, in this paper, the
computational complexity is measured by the number of arithmetic operations
(additions, multiplications, comparisons, etc.).

%

\section{Preliminaries}

\subsection{Sampling Decoding}
Consider the decoding of an $n \times n$ real-valued system. The
extension to the complex-valued system is straightforward
\cite{CongRandom}. Let $\mathbf{x}$ denote the transmitted signal
taken from a constellation $\mathcal{X}^n \subseteq \mathbb{Z}^n$.
The corresponding received signal $\mathbf{y}$ is given by
\begin{equation}
\mathbf{y}=\mathbf{H}\mathbf{x}+\mathbf{n}
\label{eqn:System Model}
\end{equation}
where $\mathbf{H}$ is an $n\times n$ full column-rank matrix of
channel coefficients and $\mathbf{n}$ is the noise vector with zero
mean and variance $\sigma^{2}$.

Given the model in (\ref{eqn:System Model}), ML decoding is shown as
follows:
\begin{equation}
\widehat{\mathbf{x}}=\underset{\mathbf{x}\in \mathcal{X}^{n}}{\operatorname{arg~min}} \, \|\mathbf{y}-\mathbf{H}\mathbf{x}\|^2
\label{eqn:ML Decoding}
\end{equation}
where $\| \cdot \|$ denotes Euclidean norm. Vector
$\mathbf{H}\mathbf{x}$ can be viewed as a lattice point of the lattice $\mathcal{L}(\mathbf{H})$ and
ML decoding corresponds to solving the CVP in the
lattice $\mathcal{L}(\mathbf{H})$. In practice, ML decoding is
always performed by sphere decoding. Due to the exponential
complexity of sphere decoding, lattice-reduction-aided decoding
is often preferred due to its acceptable complexity.

In SIC decoding (also known as Babai's nearest plane algorithm), after QR-decomposition
of the channel matrix $\mathbf{H}=\mathbf{QR}$, the system model in
(\ref{eqn:System Model}) becomes
\begin{equation}
\mathbf{y}{'}=\mathbf{Q}^T\mathbf{y}=\mathbf{R}\mathbf{x}+\mathbf{n}{'}
\label{eqn:QR}
\end{equation}
where $\mathbf{Q}$ is an orthogonal matrix and $\mathbf{R}$ is an
upper triangular matrix. At each decoding level $i=n,n-1,\ldots,1$, the pre-detection
signal $\widetilde{x}_i$ is calculated as
\begin{equation}
\widetilde{x}_i=\frac{y'_i-\sum^n_{j=i+1}r_{i,j}\widehat{x}_j}{r_{i,i}}
\label{eqn:noise effection in determination}
\end{equation}
where the decision $\widehat{x}_i$ is obtained by
rounding $\widetilde{x}_i$ to the nearest integer as
\begin{equation}
\widehat{x}_i=\lceil\widetilde{x}_i \rfloor.
\end{equation}

\begin{figure}
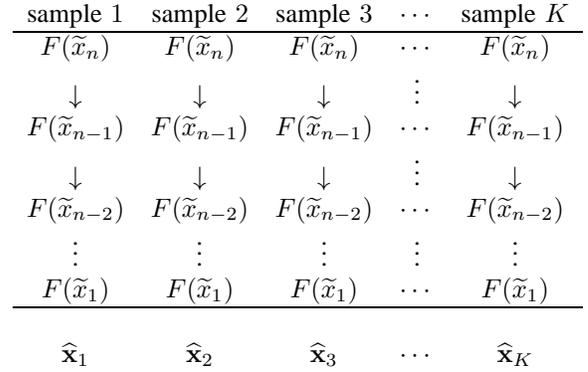

\begin{equation}
\begin{array}{ccccc}
                                 \text{sample}\,\,1 &  \text{sample}\,\,2  & \text{sample}\,\,3 & \cdots & \text{sample}\,\,K \\ \hline
                                   F(\widetilde{x}_{n}) &    F(\widetilde{x}_{n})  &  F(\widetilde{x}_{n}) & \cdots &  F(\widetilde{x}_{n})\\
                                   \downarrow & \downarrow  & \downarrow & \vdots & \downarrow\\
                                 F(\widetilde{x}_{{n}-1}) &  F(\widetilde{x}_{{n}-1})  & F(\widetilde{x}_{{n}-1})& \cdots & F(\widetilde{x}_{{n}-1}) \\
                                 \downarrow & \downarrow  & \downarrow & \vdots & \downarrow\\
                                 F(\widetilde{x}_{{n}-2}) &  F(\widetilde{x}_{{n}-2})  & F(\widetilde{x}_{{n}-2}) & \cdots & F(\widetilde{x}_{{n}-2}) \\
                                 \vdots & \vdots & \vdots & \vdots & \vdots\\
                                 F(\widetilde{x}_1) &   F(\widetilde{x}_1)  & F(\widetilde{x}_1) & \cdots & F(\widetilde{x}_1) \\  \hline
                   \\ \widehat{\mathbf{x}}_1 &   \widehat{\mathbf{x}}_2  & \widehat{\mathbf{x}}_{3} & \cdots & \widehat{\mathbf{x}}_K\\
                                \end{array} \notag
\end{equation}
  \caption{Sampling procedures in randomized sampling decoding.}
  \label{structure of randomized lattice decoding}
\end{figure}

Different from SIC decoding, in randomized sampling decoding
\cite{CongRandom}, $\widehat{x}_i$ is generated randomly from the 2$N$-integer set $\mathcal{I}$:
$\{\lfloor\widetilde{x}_i\rfloor-N+1,\ldots,\lfloor\widetilde{x}_i\rfloor,\ldots,\lfloor\widetilde{x}_i\rfloor+N\}$ centered at $\widetilde{x}_i$:
\begin{equation}
\widehat{x}_i=F(\widetilde{x}_i)
\end{equation}
where function $F(\widetilde{x}_i)$ denotes the random rounding.
\begin{figure*}[t]
\includegraphics[width=7in]{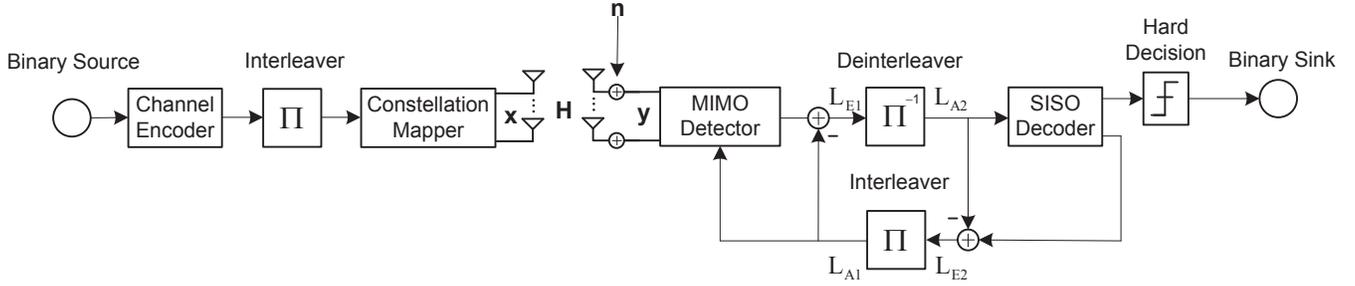}
\caption{BICM transmitter and IDD receiver in an  MIMO system.}
\label{IDD figure}
\end{figure*}
Based on Klein's sampling algorithm \cite{Klein}, the probability
of $F(\widetilde{x}_i)$ returning an integer $\widehat{x}_i^j$ ($1\leq j\leq 2N$) from the 2$N$-integer set is
calculated from the following discrete Gaussian distribution
\begin{equation}
\begin{split}
P(\widehat{x}^j_i)&=\frac{e^{-c_i(\widetilde{x}_i-\widehat{x}^j_i)^2}}{s},
\end{split}
\label{eqn:Klein}
\end{equation}
where
\begin{equation}
\begin{split}
s=s(c_i)&=\sum_{\widehat{x}^l_i\in\mathcal{I}}e^{-c_i(\widetilde{x}_i-\widehat{x}^l_i)^2}
\end{split}
\end{equation}
and $c_i=Ar^2_{i,i}$.
Note that $P(\widehat{x}^j_i) $ is a conditional probability because
the selection of previous entries are also taken into account to calculate $\widetilde{x}_i$.
As for the selection of the parameter $A$ which affects the variance of sampling probabilities,
Klein chose $A=\text{log}n/\text{min}_ir_{i,i}^2$
and \cite{CongRandom} gave a better parameter for randomized sampling
algorithm as $A=\text{log}\rho/\text{min}_ir_{i,i}^2$ where the parameter $\rho>1$
related with the sample size $K$ follows
\begin{equation}
K=(e\rho)^{2n/\rho}.
\label{KKK}
\end{equation}

The decisions $\widehat{x}_i$'s are generated level by level, and a
candidate lattice point $\widehat{\mathbf{x}}$ is obtained if all
the entries are generated. It has been demonstrated in \cite{CongRandom}
that given ${\mathbf{y}}$, the probability of a vector $\mathbf{\hat{x}}$
being sampled (also known as the sampling probability of $\mathbf{\hat{x}}$)
is lower bounded by
\begin{equation}
P(\mathbf{\hat{x}})\geq\frac{1}{\prod^n_{i=1}s(Ar_{i,i}^2)}e^{-A\|{\mathbf{y}}-\mathbf{H\hat{x}}\|^2}.
\label{sampling radius bound}
\end{equation}
By repeating this sampling procedure for
$K$ times, a candidate list of $K$ lattice points is obtained as
shown in Fig. \ref{structure of randomized lattice decoding} and
the closest one in Euclidean norm is chosen as the decoding output.

However, because samplings are random and because the $K$ samples are
independent of each other, lattice points are sampled following the
probability $P=1-(1-P(\mathbf{x}))^{K}$, which results in two inherent
problems in random sampling. On one hand, inevitable sample repetitions
in the final candidate list means unnecessary complexity is incurred.
Meanwhile, the performance is also degraded by the existence of repetitions
since most of samplings are employed to sample those lattice points with
large sampling probabilities. On the other hand, lattice points have to
face the risk of being missed during the sampling, especially for
those with small sampling probabilities on the early decoding levels,
leading to inevitable performance loss. Actually, to make sure lattice
points with a reasonable probability to be sampled, one has to increase the sample size $K$,
which leads to more sampling repetitions. Therefore, the efficiency and
performance of the randomized sampling are greatly suffered from the
randomization.

\subsection{Soft-Output Decoding}
In order to achieve near-capacity performance with low complexity
in MIMO-BICM systems, iterative detection and decoding (IDD)
proposed in \cite{Hochwald} has attracted much attention recently,
which improves the performance by iteratively exchanging the
extrinsic information between MIMO detector and soft-in soft-out
(SISO) decoder.

As shown in Fig. \ref{IDD figure}, the extrinsic information $L_{E1}$ is
calculated by the MIMO detector based on the channel observation $\mathbf{y}$
and \emph{a priori} information (API) $L_{A1}$ of the transmitted bits which
is provided by the SISO decoder. Then $L_{E1}$ is passed
through the deinterleaver to become API $L_{A2}$ to the SISO decoder,
which computes the new extrinsic information $L_{E2}$ to feed back to the
MIMO detector. Specifically, the extrinsic
information in soft-output decoding is always calculated through the
computation of the \emph{posterior} LLR for each information bit
associated with the transmitted signal $\mathbf{x}$,  which is given as
\begin{equation}
L(b_i|\mathbf{y})=\text{log}\frac{P(b_i=1|\mathbf{y})}{P(b_i=0|\mathbf{y})}
\label{eqn:LLR original}
\end{equation}
where $b_i$ is the $i$-th information bit in $\mathbf{x}$, $1\leq i \leq mn$.
Here, $m$ represents the number of bits per constellation symbol and $\mathbf{x}$
contains $mn$ information bits in all. Through the exchange of extrinsic information
in each iteration, the performance of soft-output decoding improves
gradually and we have
\begin{eqnarray}
L(b_i|\mathbf{y}) \ \ \ \ \ \ \ \ \ \ \ \ \ \ \ \ \ \ \ \ \ \ \ \ \ \ \ \ \ \ \ \ \ \ \ \ \ \ \ \ \ \ \ \ \ \ \ \ \ \ \ \ \ \ \ \ \notag \\
=\hspace{-0.2em}L_\textrm{A}(b_i)\hspace{-0.2em}+\hspace{-0.2em}\text{log}\frac{\sum_{\mathbf{x}:b_{i}=1}P(\mathbf{y}|\mathbf{x})\hspace{-0.1em}\cdot\hspace{-0.1em} \text{exp}\sum_{j\in \mathcal{J}_i}L_\textrm{A}(b_j)}{\sum_{\mathbf{x}:b_{i}=0}P(\mathbf{y}|\mathbf{x})\hspace{-0.1em}\cdot\hspace{-0.1em} \text{exp}\sum_{j\in \mathcal{J}_i}L_\textrm{A}(b_j)} \hspace{-0.3em}
\end{eqnarray}
where $L_\textrm{A}(b_i)$ denotes API of each transmitted bit
in $\mathbf{x}$
\begin{equation}
L_\textrm{A}(b_i)=\text{log}\frac{P(b_i=1)}{P(b_i=0)}
\end{equation}
and $\mathcal{J}_i$ is the set of indices $j$ with
\begin{equation}
\mathcal{J}_i=\{j|j=1,\ldots, mn, j\neq i\}.
\end{equation}

In the absence of API, we suppose all the bits in $\mathbf{x}$
have the same probability to be 0 or 1 before $\mathbf{y}$ is observed
as $P(b_i=1)=P(b_i=0)=\frac{1}{2}$.
Then, for simplicity, the $L$-value in (\ref{eqn:LLR original}) becomes
\cite{Hochwald,LarssonPartialMarginalization}
\begin{equation}
L(b_i|\mathbf{y})=\text{log}\frac{\sum_{\mathbf{x}:b_{i}=1}\text{exp}\ (-\frac{1}{2\sigma^2}\parallel \mathbf{y}-\mathbf{H}\mathbf{x} \parallel^2)}{\sum_{\mathbf{x}:b_{i}=0}\text{exp}\ (-\frac{1}{2\sigma^2}\parallel \mathbf{y}-\mathbf{H}\mathbf{x} \parallel^2)}.
\label{eqn:LLR}
\end{equation}

The straightforward way to calculate the $L$-value in (\ref{eqn:LLR}) is MAP
algorithm which computes the sums that contain $2^{mn}$ terms. Due
to the exponentially increased complexity of MAP, one has to resort to
approximations to reduce the complexity.

As one of the approximation scenarios, Max-Log approximation tries to
approximate the sums in (\ref{eqn:LLR}) only with their largest terms
\cite{Renqiuwangsoft,Studersoft}:
\begin{equation}
L(b_i|\mathbf{y})\approx\text{log}\frac{\text{max}_{\mathbf{x}:b_{i}=1}\text{exp}\ (-\frac{1}{2\sigma^2}\parallel \mathbf{y}-\mathbf{H}\mathbf{x} \parallel^2)}{\text{max}_{\mathbf{x}:b_{i}=0}\text{exp}\ (-\frac{1}{2\sigma^2}\parallel \mathbf{y}-\mathbf{H}\mathbf{x} \parallel^2)}.
\label{eqn:LLR Max Log}
\end{equation}

However, to obtain the largest terms in (\ref{eqn:LLR Max Log}), sphere
decoding is applied, which incurs exponential increment complexity.
To achieve a polynomial complexity, suboptimal hard decoding schemes like
SIC are used to solve those maximization problems approximately.
Unfortunately, the decoder performance is poor even under the help
of the lattice reduction technique.

As for another approximation schemes, list sphere decoding (LSD)
proposed in \cite{Hochwald} restricts the sums in (\ref{eqn:LLR}) into a
much smaller size, which uses sphere decoding with a certain radius to
perform the candidate admission as follows
\begin{equation}
\mathcal{S}\triangleq\{\mathbf{x}\in\mathcal{X}^n:\|\mathbf{y}-\mathbf{Hx}\|^2\leq R^2\}.
\label{eqn:definition of S}
\end{equation}
Here, $R$ denotes the sphere radius and a larger $R$ means
a better approximation and also, a higher computational complexity.
Then, the calculation of \emph{L}-value in (\ref{eqn:LLR}) can be
written as
\begin{equation}
L(b_i|\mathbf{y})\approx\text{log}\frac{\sum_{\mathbf{x}\in\mathcal{S}:b_{i}=1}\text{exp}\ (-\frac{1}{2\sigma^2}\parallel \mathbf{y}-\mathbf{H}\mathbf{x} \parallel^2)}{\sum_{\mathbf{x}\in\mathcal{S}:b_{i}=0}\text{exp}\ (-\frac{1}{2\sigma^2}\parallel \mathbf{y}-\mathbf{H}\mathbf{x} \parallel^2)}.
\label{eqn:LLR LSD}
\end{equation}

Although lattice reduction technique can be applied to reduce its
complexity, LSD schemes still suffer from a high complexity cost due
to the application of sphere decoding. In particular, unlike finding the closest
lattice point in lattice decoding, sphere decoding in LSD tries to
enumerate all the lattice points within a constant sphere radius
$R$. Additionally, since the selection of the sphere
radius $R$ affects the performance of LSD, there is still an open
question about the upper bound of $R$ when LSD achieves near-MAP
performance.

\section{Derandomized Sampling Decoding}

In this section, we propose a derandomized sampling algorithm to
solve the afore-mentioned problems in randomized sampling decoding, namely,
repetition and missing of certain lattice points. Specifically,
the sampling procedure of the derandomized sampling algorithm is performed
level by level with $i=n, n-1, \ldots, 1$ as follows:

\begin{algorithm}
\caption{Derandomized sampling on decoding level $i$}
\begin{algorithmic}
\State Compute $\widetilde{x}_i=\frac{y'_i-\sum^n_{l=i+1}r_{i,l}\widehat{x}_l}{r_{i,i}}$
\For {$j=$1,2,\ldots,2$N$}
    \State Compute $P(\widehat{x}^j_i)=\frac{e^{-c_i(\widetilde{x}_i-\widehat{x}^j_i)^2}}{s}$
    \State Compute $E(\widehat{x}^j_i)=\lceil K_{i+1}P(\widehat{x}^j_i)\rfloor$
    \If {$E(\widehat{x}^j_i)<1$}
    \State $\widehat{x}^j_i$ is ignored, sampling based on $\widehat{x}^j_i$ terminates
    \ElsIf {$E(\widehat{x}^j_i)=1$}
    \State Let $\widehat{x}_i=\widehat{x}^j_i$, SIC is performed based on the detected
    \State entries $\widehat{x}_n,\ldots,\widehat{x}_i$ to return a candidate lattice point
    \ElsIf {$E(\widehat{x}^j_i)>1$}
          \If {$i>1$}
          \State Let $\widehat{x}_i=\widehat{x}^j_i$, $K_{i}=K_{i+1}P(\widehat{x}^j_i)$, start derandomi-
          \State zed sampling from the next level $i-1$ based on
          \State the detected entries $\widehat{x}_n,\ldots,\widehat{x}_i$
          \ElsIf {$i=1$}
          \State Let $\widehat{x}_1=\widehat{x}^j_1$, return $\widehat{x}_n,\ldots,\widehat{x}_1$ as a candidate
          \State lattice point
          \EndIf
    \EndIf
\EndFor

\end{algorithmic}
\end{algorithm}

At decoding level $i$, sample size $K_{i+1}$ is allocated to candidate integers
according to $K_{i}= K_{i+1}P(\widehat{x}^j_i)$ and all the integers with $E(\widehat{x}^j_i)\geq 1$
are deterministically sampled. Note that $K_{i}$ is not necessarily an
integer any more. For integers with $E(\widehat{x}^j_i)>1$, after
updating the size $K_{i}$, sampling continues from the next level in
the same way. Note that when $K=1$, derandomized sampling decoding
performs the same with SIC decoding by always selecting the integer with the largest
probability. Hence, for integers with $E(\widehat{x}^j_i)=1$, SIC is applied directly
to obtain a candidate lattice point. Finally, among all the candidate lattice points,
the closest one is selected as the solution. By performing the sampling based on
the threshold $E(\widehat{x}^j_i)\geq1$ at each decoding level, the whole
sampling process becomes deterministic. The risk of lattice points being missed
during the sampling is greatly reduced, which means the probability of sampling the
closest lattice point is improved.
\begin{figure}
\includegraphics[width=3.6in]{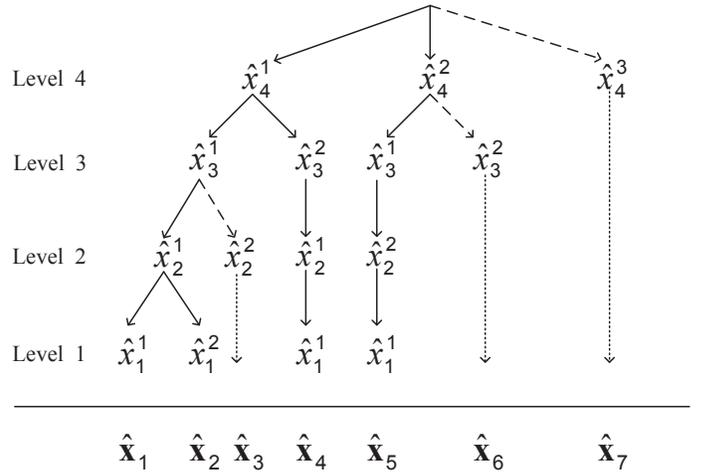}\\
\caption{Illustration of the tree structure in derandomized sampling decoding with ${n}=4$. The solid and dashed lines represent the corresponding candidate integer being sampled by $E(\widehat{x}^j_i)>1$ and $E(\widehat{x}^j_i)=1$ respectively. The dotted line denotes performing SIC detection in the rest of levels due to $E(\widehat{x}^j_i)=1$.}
\label{structure of derandomized lattice decoding}
\end{figure}

Unlike the parallel structure of the random sampling, derandomized sampling decoding
admits a tree structure as shown in Fig. \ref{structure of derandomized lattice decoding}.
The final candidate list is generated by traversing the tree from level $n$ to
level 1 rather than by $K$ independent paths. From this perspective, derandomized
sampling functions like a pruning algorithm in sphere decoding \cite{HassibiPruning,ZhaoSD,ShimSD}
which prunes branches $E(\widehat{x}^j_i)<1$. Thanks to the tree structure, there are no
sampling repetitions during the whole sampling process while necessary calculations of
sampling probabilities $P(\widehat{x}^j_i)$  in branches $E(\widehat{x}^j_i)>1$ are
performed only once, saving a lot of complexity. Therefore, different from randomized
sampling decoding and other decoding schemes establishing a candidate list with a constant
size around the SIC output \cite{MaxiaoliSoft,Shimokawa}, the size of
the final candidate list $K_{\text{final}}\leq K$ is variable, which
means the sample size $K$ set initially in the derandomized sampling algorithm is
actually a \textit{nominal sample size} of the candidate list.

As a \textit{nominal sample size}, $K$ is essentially a parameter in the
threshold $E(\widehat{x}^j_i)=\lceil K_{i+1}P(\widehat{x}^j_i)\rfloor\geq1$ used to evaluate the sampling
performance. With the increment of $K$, the complexity improves
gradually since more lattice points will be sampled. Note that
$K$ is not the real size of the final sampling list,
the complexity of derandomized sampling decoding in fact grows
slowly with its increment. Because problems
caused by randomization are overcome, derandomized algorithm achieves
desirable improvement in both performance and complexity.

\subsection{Algorithm Analysis}
The operation of the derandomized sampling algorithm relies on the
notion of the sampling probability which is calculated by (\ref{eqn:Klein}).
According to the defined threshold $E(\widehat{x}^j_i)\geq1$, at each decoding level,
an integer candidate $\widehat{x}_i^j$ for the entry $x_i$ will be sampled
if and only if
\begin{equation}
K_{i+1}P(\widehat{x}_i^j) \geq \frac{1}{2}.
\label{eqn:the defined threshold}
\end{equation}
Note that the sampling probability $P(\widehat{x}_i^j)$ calculated by (\ref{eqn:Klein})
is a conditional probability based on the entries of previous levels.
As sampling is performed from level $n$ to level 1, the sampling probability
$P(\mathbf{x})$ about lattice point $\mathbf{x}$ is essentially the product of
its entries' sampling probability, which is lower bounded by (\ref{sampling radius bound}):
\begin{equation}
P(\mathbf{x})=\prod_{i=1}^nP(\widehat{x}_{n+1-i}^j)\geq\frac{1}{\prod^n_{i=1}s(Ar_{i,i}^2)}e^{-A\|{\mathbf{y}}-\mathbf{Hx}\|^2}.
\label{probability Px}
\end{equation}

\newtheorem{my}{Proposition}
\begin{my}
Given the nominal sample size $K$, lattice points with sampling probability
\begin{equation}
P(\mathbf{x})\geq \frac{1}{2K}
\label{eqn:probability of derandomized}
\end{equation}
will be deterministically sampled by derandomized sampling algorithm.
\end{my}

\emph{Proof:} Consider sampling an $n$-dimensional lattice point $\mathbf{x}$ by derandomized sampling
algorithm. Obviously, with the initial sample size $K$, its first entry $\widehat{x}_n^j$ on
level $n$ will be sampled if
\begin{equation}
KP(\widehat{x}_n^j)\geq\frac{1}{2}.
\end{equation}

Based on the selection of $\widehat{x}_n^j$, its updated sample size $K_n$ on the next level is
calculated as
\begin{equation}
K_n=KP(\widehat{x}_n^j).
\end{equation}
Then, on level $n-1$, the first two entries
of $\mathbf{x}$ will be obtained when
\begin{equation}
K_nP(\widehat{x}_{n-1}^j)=KP(\widehat{x}_n^j)P(\widehat{x}_{n-1}^j)\geq\frac{1}{2}.
\end{equation}

By induction, $\mathbf{x}$ will be deterministically sampled if the following
condition holds
\begin{equation}
K\prod_{i=1}^nP(\widehat{x}_{n+1-i}^j)=KP(\mathbf{x})\geq\frac{1}{2}.
\end{equation}
Thus, the conclusion follows, completing the proof.

As for the randomized sampling in \cite{CongRandom}, because the $K$ times sampling is independent
of each other, the probability of missing $\mathbf{x}$ is calculated as
$(1-P(\mathbf{x}))^K$, which means one has to increase the sample size $K$ to
ensure $\mathbf{x}$ a high probability of being sampled. In particular, given
sample size $K$, lattice points with sampling probability
\begin{equation}
P(\mathbf{x})\geq \frac{1}{K}
\label{eqn:probability of random}
\end{equation}
will be found by randomized sampling algorithm with probability $P\geq 1- 1/e$. Through the comparison between (\ref{eqn:probability of derandomized}) and (\ref{eqn:probability of random}), to sample the same lattice point $\mathbf{x}$, the required sample size of the derandomized
sampling algorithm is less than the half of that in randomized sampling algorithm:
\begin{equation}
K_{\text{Derandomized}}<\frac{1}{2}K_{\text{Random}}.
\end{equation}
On the other hand, with the same sample size $K_{\text{Derandomized}}=K_{\text{Random}}$,
derandomized sampling algorithm has the ability to obtain more lattice points than
randomized sampling, which brings further performance improvement. Since
$K_{\text{Derandomized}}$ is the nominal sample size, for the same sample size
derandomized sampling algorithm still achieves much lower complexity than randomized
sampling. More precisely, when $K=1$, the complexity of the derandomized sampling
algorithm is $O(n^2)$ by invoking the calculation of the sampling probability in
(\ref{eqn:Klein}) for $n$ times. For $K>1$, as computations in sampling procedures
are reduced by removing all the repetitions, the number of recalling the calculation in
(\ref{eqn:Klein}) is much less than $Kn$, which means the complexity is much smaller
than $K\cdot O(n^2)$. Due to the uncertainty in this procedure, it is preferable
to denote the complexity of the derandomized sampling algorithm by $K\cdot O(n^2)$, which means
a polynomial complexity with respect to the dimension $n$. Obviously,
without suffering from the effect of the randomization, derandomized sampling algorithm shows
great potential in both performance and complexity.

\subsection{Optimization of the Parameter $A$}
As a parameter which controls the variance of sampling probabilities, parameter $A$
has a significant impact on the final decoding performance. Due to the consideration of complexity, the initial sample size $K$ of sampling algorithms is always
limited, which means finding the optimum $A$ to exploit the sampling potential for a given
$K$ is the key. In order to determine the optimum choice of $A$ in
the derandomized sampling algorithm, let $A=\text{log}\rho/\text{min}_ir_{i,i}^2$ where $\rho>1$,
then $\rho$ becomes the parameter needed to be optimized.

It has been demonstrated in \cite{Klein} that
\begin{equation}
\overset{n}{\underset{i=1}{\prod}}s(c_i)\leq e^{\frac{2n}{\rho}(1+O(\rho^{-3}))}.
\label{rho1}
\end{equation}
Because $\rho>1$, the term $O(\rho^{-3})$ in (\ref{rho1}) will be negligible if
$\rho$ is sufficiently large. Assume $\rho$ satisfies this weak condition, the
sampling probability of $\mathbf{x}$ shown in (\ref{probability Px}), which
is calculated based on the discrete Gaussian distribution, can be further derived as follows
\begin{equation}
P(\mathbf{x})\geq e^{-\frac{2n}{\rho}}\cdot \rho^{-\|\mathbf{y}-\mathbf{H}\mathbf{x}\|^2/\text{min}_ir_{i,i}^2}.
\label{distance1}
\end{equation}

Since lattice points with $P(\mathbf{x})\geq\frac{1}{2K}$ will be deterministically
sampled by derandomized sampling algorithm, motivated by (\ref{distance1}), let
\begin{equation}
e^{-\frac{2n}{\rho}}\cdot\rho^{-\|\mathbf{y}-\mathbf{H}\mathbf{x}\|^2/\text{min}_ir_{i,i}^2}\geq\frac{1}{2K},
\end{equation}
and we have
\begin{equation}
\|\mathbf{y}-\mathbf{H}\mathbf{x}\|\leq \text{min}_ir_{i,i}\cdot\sqrt{\text{log}_\rho(2Ke^{-2n/\rho})},
\label{distance2}
\end{equation}
which means lattice points with $\|\mathbf{y}-\mathbf{H}\mathbf{x}\|$ less than
the right-hand side (RHS) of (\ref{distance2}) must be obtained.

In order to exploit the potential of the derandomized sampling algorithm for the
best decoding performance, parameter $\rho$ is selected carefully to maximize the
upper bound shown in (\ref{distance2}). Therefore, let the derivative about
$\text{log}_\rho(2Ke^{-2n/\rho})$ versus $\rho$ be zero, the optimum $\rho_{\text{o}}$
given sample size $K$ in the derandomized sampling algorithm can be finally determined as follows
\begin{equation}
K=\frac{1}{2}(e\rho_{\text{o}})^{2n/\rho_{\text{o}}}.
\label{relationship K and rho}
\end{equation}

Obviously, the optimum $\rho_{\text{o}}$ for the randomized sampling algorithm shown in
(\ref{KKK}) is not the optimum solution in the derandomized sampling algorithm.
According to (\ref{relationship K and rho}), it is easy to check that the parameter
$\rho_{\text{o}}$ monotonically decreases with respect to the increment of the sample
size $K$.

In the view of lattice decoding, derandomized sampling algorithm
will give the closest lattice point if the distance between $\mathbf{y}$
and lattice $\mathcal{L}(\mathbf{H})$ is less than the
RHS of (\ref{distance2}). Therefore, the RHS of (\ref{distance2}) can
be regarded as the decoding radius in the notion of bounded distance
decoding (BDD). By substituting (\ref{relationship K and rho}) into
(\ref{distance2}), the optimum decoding radius $R_{\text{Derandomized}}$
of the derandomized sampling algorithm is derived as
\begin{equation}
R_{\text{Derandomized}} \triangleq \sqrt{\frac{2n}{\rho_{\text{o}}}}\ \text{min}_ir_{i,i}.
\label{optimum decoding radius}
\end{equation}

Clearly, with the increment of $K$, $R_{\text{Derandomized}}$ improves
gradually resulting in better decoding performance. We need to
emphasize that the decoding radius $R_{\text{Derandomized}}$ may
be much larger than the value defined in (\ref{optimum decoding radius})
because the derivation is only based on the lower bound of
$P(\mathbf{x})$ shown in (\ref{distance1}).

\subsection{Upper Bound on the Sample Size $K$}
We now give an explicit value of $K$ when derandomized sampling
decoding achieves near-ML performance. To do this, the total probability of
samples in the final candidate list is derived based on the truncation of the discrete Gaussian
distribution (\ref{eqn:Klein}).

As shown in \cite{CongRandom}, the probability that the integer
$\widehat{x}_i$ generated by random rounding $F(\widetilde{x}_i)$ is
located within the 2\emph{N} integers around $\widetilde{x}_i$ is
bounded by
\begin{equation}
P_{2N}\geq1-O(\rho^{-N^2}).
\end{equation}

Because $\rho>1$, the term $O(\rho^{-N^2})$ decays exponentially,
meaning a finite truncation with moderate $N$ achieves an accurate
approximation. Normally, 3-integer approximation is sufficient:
\begin{equation}
P(\widehat{x}_i^1) + P(\widehat{x}_i^2) + P(\widehat{x}_i^3) \approx
1. \label{eqn:3-integer approximation}
\end{equation}

Since these probabilities follow the discrete Gaussian distribution,
they decrease monotonically with the distance from
$\widetilde{x}_i$. Let us order them as follows
\begin{equation}
P(\widehat{x}_i^1) \geq P(\widehat{x}_i^2) \geq P(\widehat{x}_i^3).
\label{eqn:Relationship}
\end{equation}

As shown in (\ref{eqn:noise effection in determination}),
$\widetilde{x}_i$ is subject to the effect of noise. Intuitively,
$\widetilde{x}_i$ tends to be close to an integer for small noise
while it tends to be halfway between two integers for large noise.
Since $\widetilde{x}_i$ is the peak of the continuous Gaussian
distribution associated with the discrete one (\ref{eqn:Klein}), we
define the \textit{worst case} in sampling as the one where
$\widetilde{x}_i$ is centered between two integers.

Because the random noise makes it hard for an exact analysis, we only
consider the worst-case scenario in samplings. Then, under the
3-integer approximation in (\ref{eqn:3-integer approximation}), the
following holds in the worst case:
\begin{equation}
P(\widehat{x}_i^1) = P(\widehat{x}_i^2) \gg P(\widehat{x}_i^3)
\label{eqn:probabilities relationship in worst case}
\end{equation}
where $P(\widehat{x}_i^3)$ is much smaller due to exponential decay
of the probability with the distance.

Now, let us calculate the total probability of lattice points
sampled by derandomized sampling, in the worst case.  Consider
the level $n$ first. Obviously, according to (\ref{eqn:the defined threshold}),
the first two integers will be sampled if $K>2$. If $P(\widehat{x}^3_n)\geq\frac{1}{2K}$,
all the 3 integers around $\widetilde{x}_n$ are deterministically
sampled. On the other hand, if $P(\widehat{x}^3_n)<\frac{1}{2K}$,
integer $\widehat{x}^3_n$ will be discarded while the summation of
probabilities of the other two integers will be larger than
$1-\frac{1}{2K}$ according to (\ref{eqn:3-integer approximation}).
Therefore, given the nominal sample size $K$, the sum probability of
samples on the level $n$ is bounded by
\begin{equation}
P(\text{level}\,n)\geq1-\frac{1}{2K}.
\end{equation}

To further derive the lower bound of the total probability of
samples, we assume the third sample $\widehat{x}^3_i$ at
the each sampling level is always discarded. Then, still in
the worst case, the total probability of samples on the
level $n-1$ is given by
\begin{eqnarray}
&&P(\text{level}\,n-1)\hspace{-0.3em}\geq \hspace{-0.3em}(1\hspace{-0.3em}-\hspace{-0.3em}\frac{1}{2K})\hspace{-0.3em}\left[\frac{1}{2}\hspace{-0.3em}\left(\hspace{-0.3em}1\hspace{-0.3em}-\hspace{-0.3em}\frac{1}{K}\hspace{-0.3em}\right)\hspace{-0.3em}+\hspace{-0.3em}\frac{1}{2}\hspace{-0.3em}\left(\hspace{-0.3em}1\hspace{-0.3em}-\hspace{-0.3em}\frac{1}{K}\hspace{-0.3em}\right)\right] \notag\\
&&\,\,\,\,\,\,\,\,\,\,\,\,\,\,\,\,\,\,\,\,\,\,\,\,\,\,\,\,\,\,\,\,\,\,\,\,=(1-\frac{1}{2K})(1-\frac{1}{K}).
\end{eqnarray}

Similarly, on the level $n-p$, the total probability of samples in
the worst case can be lower bounded by
\begin{equation}
P(\text{level}\,n-p)> \prod_{i=1}^p(1-\frac{2^{i-2}}{K}).
\label{eqn:p level probability}
\end{equation}

Therefore, the total probability of sampled lattice points in the
derandomized sampling algorithm is lower bounded by a function of $K$. We
define a parameter $\eta$ to evaluate the decoding performance as
\begin{equation}
P_{\textrm{total}}=P(\text{level}\,1) > \eta \triangleq
\prod_{i=1}^n(1-\frac{2^{i-2}}{K}). \label{lower bound y}
\end{equation}

Obviously, the lower bound $\eta$ increases with $K$ and a larger
$\eta$ means a higher probability of the closest lattice point being
sampled. Thus, derandomized sampling decoding can be used to
approximate ML decoding as $\eta$ approaches 1.

The lower bound (\ref{lower bound y}) is loose because it
quantifies the probability in the worst case. For $\eta$ close to 1,
$K$ can be very huge (in fact exponential). A lower bound in the
average case is an open question. Because the noise is random, the
average-case probability may be more useful.

In order to obtain a better estimate, the idea of the fixed-complexity
sphere decoding (FSD), which also follows a tree structure in
decoding, is exploited. Different from the standard sphere decoding, it
only performs the full search in the upper $p$ levels known as the
full-expansion stage while SIC is applied on the rest of
levels. It has been proved in \cite{JaldenFSD} that by applying the
channel matrix ordering to make sure signals with the largest
postprocessing noise amplification are detected in the
full-expansion stage, FSD algorithm yields near-ML performance in
high SNR if it satisfies:
\begin{equation}
(p+1)^2\geq n
\label{full diversity level}
\end{equation}
where $p$ is the number of levels in the full-expansion stage.

We propose to use sampling in the full-expansion stage of FSD. With the
suitable channel matrix ordering, the modified sampling decoder also
consists of two stages. Candidate values on the upper $p$ levels are
sampled based on the lower bound $\eta$ while decodings on the
remaining levels are processed by SIC (i.e., derandomized sampling
decoding with $K=1$).

According to (\ref{eqn:p level probability}) and (\ref{full
diversity level}), if we set $\eta$ to a value near 1 on the upper
$p$ decoding levels, then the decoder will achieve near-ML
performance:
\begin{equation}
P_{\textrm{total}}=P(\text{level}\,n-p)>
\eta=\prod_{i=1}^{p}(1-\frac{2^{i-2}}{K}). \label{new lower bound}
\end{equation}

Compared with (\ref{lower bound y}), the lower bound (\ref{new lower
bound}) is better because $p$ is much smaller than $n$ meaning
the value of $K$ achieving the same $\eta$ is greatly reduced.
Here, we define $\eta=0.9$ representing near-ML performance. Then,
according to (\ref{new lower bound}), the corresponding $K$ who denotes
the upper bound of the sample size in derandomized sampling decoding can be
easily calculated.

Note that the derandomized sampling algorithm with $K=1$ performs the same
with SIC in lattice decoding. Thus, the decoder enjoys flexible performance between SIC and near-ML
by adjusting $K$. Although larger $\eta>0.9$ will bring further performance
improvement, it approaches ML performance asymptotically with the
exponential increment of $K$, which is meaningless due to the consideration
of complexity.

\section{Derandomized sampling algorithm in soft-output decoding}
In this section, we show that the proposed derandomized sampling algorithm can also be used as an
efficient tool to implement soft-output decoding in MIMO systems. By
generating a list of lattice points around the ML estimate $\text{min}_{\mathbf{x}\in \mathcal{X}^n}\|\mathbf{y}-\mathbf{Hx}\|$, derandomized
sampling algorithm in soft-output decoding actually functions as an approximation
scheme like list sphere decoding (LSD) in \cite{Hochwald}. To establish the trade-off
with respect to the sample size $K$, we firstly give an upper bound of the sphere
radius $R$ in LSD.
\begin{figure*}
\begin{center}
\includegraphics[width=4.4in]{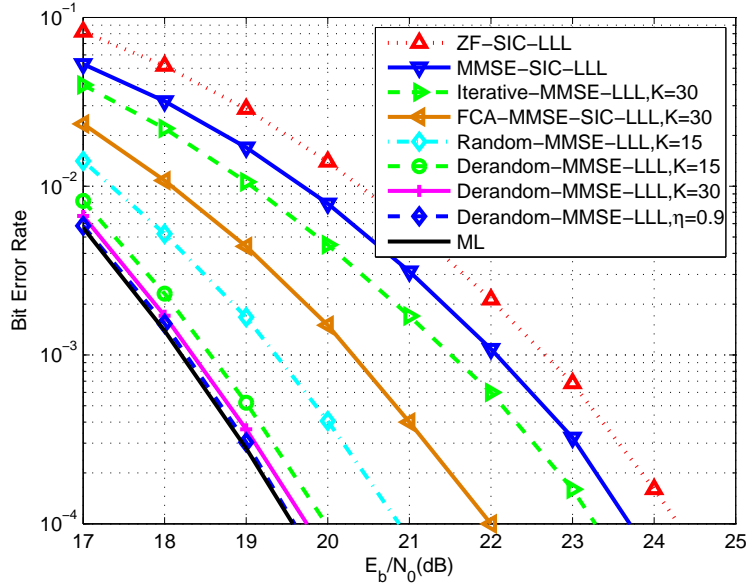}\\
\end{center}
\vspace{-1em}
\caption{Bit error rate versus average SNR per bit for the uncoded $10 \times 10$ MIMO system using 64-QAM.}
  \label{simulation 1}
\end{figure*}

\subsection{Upper Bound on the Sphere Radius $R$ in LSD}
Given $\mathbf{y}\in \mathbb{R}^n$, we define a function $f: \mathbb{R}^n
\rightarrow \mathbb{R}$ over the $n$-dimensional lattice $\mathcal{L}(\mathbf{H})$ as
\begin{equation}
f_{\mathcal{L}(\mathbf{H})}(\mathbf{y})=\frac{\sum_{\mathbf{v}\in \mathcal{L}(\mathbf{H})} \text{exp}(-\pi\|\mathbf{y}-\mathbf{v}\|^2)}{\sum_{\mathbf{v}\in \mathcal{L}(\mathbf{H})} \text{exp}(-\pi\|\mathbf{v}\|^2)},
\end{equation}
where $\mathbf{v}\in \mathbb{R}^n$ denotes the lattice point of $\mathcal{L}(\mathbf{H})$.
Thus the LLR in soft-output decoding shown in (\ref{eqn:LLR}) can be expressed by $f$-function as
\begin{eqnarray}
L(b_i\mid \mathbf{y}) &=&\text{log}\frac{\sum_{\mathbf{x}:b_{i}=1}\text{exp}(-\pi\cdot\frac{1}{2\pi\sigma^2}\parallel \mathbf{y}-\mathbf{H}\mathbf{x} \parallel^2)}{\sum_{\mathbf{x}:b_{i}=0}\text{exp}(-\pi\cdot\frac{1}{2\pi\sigma^2}\parallel \mathbf{y}-\mathbf{H}\mathbf{x} \parallel^2)}\notag\\
&=&\text{log}\frac{\frac{\sum_{\mathbf{x}:b_{i}=1}\text{exp}(-\pi\cdot\frac{1}{2\pi\sigma^2}\parallel \mathbf{y}-\mathbf{H}\mathbf{x} \parallel^2)}{\sum_{\mathbf{x}\in \mathcal{X}^n}\text{exp}(-\pi\cdot\frac{1}{2\pi\sigma^2}\parallel \mathbf{H}\mathbf{x} \parallel^2)}}{\frac{\sum_{\mathbf{x}:b_{i}=0}\text{exp}(-\pi\cdot\frac{1}{2\pi\sigma^2}\parallel \mathbf{y}-\mathbf{H}\mathbf{x} \parallel^2)}{\sum_{\mathbf{x}\in \mathcal{X}^n}\text{exp}(-\pi\cdot\frac{1}{2\pi\sigma^2}\parallel \mathbf{H}\mathbf{x} \parallel^2)}}\notag\\
&=&\text{log}\frac{f_{\mathcal{L}(\frac{1}{\sigma\sqrt{2\pi}}\mathbf{H}),\mathbf{x}:b_i=1}(\frac{1}{\sigma\sqrt{2\pi}}\mathbf{y})}{f_{\mathcal{L}(\frac{1}{\sigma\sqrt{2\pi}}\mathbf{H}), \mathbf{x}:b_i=0}(\frac{1}{\sigma\sqrt{2\pi}}\mathbf{y})}.
\label{FX}
\end{eqnarray}

Accordingly, the $L$-value computation is converted into the
calculation of function $f_{\mathcal{L}(\frac{1}{\sigma\sqrt{2\pi}}\mathbf{H})}(\frac{1}{\sigma\sqrt{2\pi}}\mathbf{y})$.
Here, the lattice point $\mathbf{v}$ in $\mathcal{L}(\mathbf{H})$ is expressed by $\mathbf{Hx}$, where $\mathbf{x}\in\mathcal{X}^n$ is an integer vector.
To further exploit the function $f$, we invoke the following lemma in \cite{Banaszczyk}.
\begin{flushleft}
\textbf{Lemma 1} (\hspace{-0.0001em}\cite{Banaszczyk})\textbf{.} \emph{For any n-dimensional lattice}
$\mathcal{L}(\mathbf{H})$, $\mathbf{y}\in \mathbb{R}^n$ \emph{and} $c>\frac{1}{\sqrt{2\pi}}$, \emph{one has}
\end{flushleft}
\begin{eqnarray}
\frac{\sum_{\mathbf{v}\in \mathcal{L}(\mathbf{H}),\parallel \mathbf{y}-\mathbf{v} \parallel>c\sqrt{n}}\text{exp}(-\pi\|\mathbf{y}-\mathbf{v}\|^2)}{\sum_{\mathbf{v}\in \mathcal{L}(\mathbf{H})}\text{exp}(-\pi\|\mathbf{v}\|^2)} \hspace{-0.3em}&\hspace{-0.3em}\hspace{-0.3em}\hspace{-0.3em}\leq\hspace{-0.3em}\hspace{-0.3em}\hspace{-0.3em}&\hspace{-0.3em} 2\left(c\sqrt{2\pi e}\cdot e^{-\pi c^2}\hspace{-0.2em}\right)^n \notag\\
\hspace{-0.3em}&\hspace{-0.3em}=\hspace{-0.3em}&\hspace{-0.3em}2^{-\Omega(n)}.
\end{eqnarray}

According to Lemma 1, we obtain
\begin{eqnarray}
f_{\mathcal{L}(\mathbf{H})}(\mathbf{y})=\frac{\sum_{\mathbf{v}\in \mathcal{L}(\mathbf{H}), \|\mathbf{y}-\mathbf{v}\|\leq c\sqrt{n}} \text{exp}(-\pi\|\mathbf{y}-\mathbf{v}\|^2)}{\sum_{\mathbf{v}\in \mathcal{L}(\mathbf{H})} \text{exp}(-\pi\|\mathbf{v}\|^2)}\ \ \ \ \ \ \ \ \ \ \ \ \ \ \ \ \ \  \notag\\
+\frac{\sum_{\mathbf{v}\in \mathcal{L}(\mathbf{H}), \|\mathbf{y}-\mathbf{v}\|> c\sqrt{n}} \text{exp}(-\pi\|\mathbf{y}-\mathbf{v}\|^2)}{\sum_{\mathbf{v}\in \mathcal{L}(\mathbf{H})} \text{exp}(-\pi\|\mathbf{v}\|^2)}\ \ \ \ \ \ \ \ \ \ \ \notag
\end{eqnarray}
\vspace{-0.3cm}
\begin{eqnarray}
\ \ \ \leq\frac{\sum_{\mathbf{v}\in \mathcal{L}(\mathbf{H}), \|\mathbf{y}-\mathbf{v}\|\leq c\sqrt{n}} \text{exp}(-\pi\|\mathbf{y}-\mathbf{v}\|^2)}{\sum_{\mathbf{v}\in \mathcal{L}(\mathbf{H})} \text{exp}(-\pi\|\mathbf{v}\|^2)}+2^{-\Omega(n)}.
\label{FX disassemble}
\end{eqnarray}

As to the second term in the RHS of (\ref{FX disassemble}), it decays
exponentially with the dimension $n$. Assume $n$ is sufficiently large, then $f_{\mathcal{L}(\mathbf{H})}(\mathbf{y})$ is dominated by the set of lattice
points within the radius $R=\sqrt{n/2\pi}$ centering at $\mathbf{y}$. Back to $f_{\mathcal{L}(\frac{1}{\sigma\sqrt{2\pi}}\mathbf{H})}(\frac{1}{\sigma\sqrt{2\pi}}\mathbf{y})$
in (\ref{FX}), correspondingly, lattice points in the corresponding set
denoted by $S$ should satisfy the following condition as
\begin{eqnarray}
\frac{1}{\sigma\sqrt{2\pi}}\|\mathbf{y}-\mathbf{v}\|&\leq&\sqrt{\frac{n}{2\pi}}
\end{eqnarray}
and we have
\begin{eqnarray}
\|\mathbf{y}-\mathbf{v}\|&\leq&\sqrt{n}\sigma.
\end{eqnarray}

Based on the fact as shown in (\ref{FX}) that
\begin{eqnarray}
f_{\mathcal{L}(\frac{1}{\sigma\sqrt{2\pi}}\mathbf{H})}(\frac{1}{\sigma\sqrt{2\pi}}\mathbf{y})\propto
\sum_{\mathbf{x}\in \mathcal{X}^{n}}\text{exp}\ (-\frac{1}{2\sigma^2}\parallel \mathbf{y}-\mathbf{H}\mathbf{x} \parallel^2),
\end{eqnarray}
the key of solving the $L$-value computation depends on lattice
points within the radius $\sqrt{n}\sigma$. In other words, it can
be interpreted as that, with the sphere radius $R=\sqrt{n}\sigma$,
LSD could achieve MAP performance within a
negligible loss by only exploiting lattice points in the set $S$
shown in (\ref{eqn:definition of S}). Thus, the sphere radius
$R$ in LSD is upper bounded by
\begin{eqnarray}
R\leq\sqrt{n}\sigma.
\label{eqn:upper bound of LSD}
\end{eqnarray}
Note that with the increase of SNR, the radius $R$ shrinks gradually
saving a lot of complexity.

\subsection{Derandomized Sampling in Soft-Output Decoding}
Given the sphere radius $R$, LSD performs sphere decoding to obtain
all the lattice points within $R$. However, it is known that sphere decoding
is impractical due to its exponentially increased complexity. Instead of enumerating
lattice points within $R$ by exhaustive search, derandomized sampling algorithm
generates lattice points by sampling from a Gaussian-like distribution, which is
more efficient than LSD due to its polynomial complexity.

As it shown in (\ref{sampling radius bound}), the lower bound of the sampling probability
resembles a Gaussian distribution over the lattice. The closer $\mathbf{Hx}$ to
$\mathbf{y}$, the larger lower bound. Therefore, lattice points closer to
$\mathbf{y}$ are more likely to be sampled due to larger sampling probability lower bounds. In this way,
the derandomized sampling algorithm could find a number of lattice points with small values of $\|\mathbf{y}-\mathbf{Hx}\|$ around the ML estimate.
By restricting the original set of sums in (\ref{eqn:LLR}) into a much smaller one denoted
by $\mathcal{C}$, the LLR calculation by derandomized sampling algorithm can be written as
\begin{equation}
L(b_i|\mathbf{y})\approx\text{log}\frac{\sum_{\mathbf{x}\in\mathcal{C}:b_{i}=1}\text{exp}\ (-\frac{1}{2\sigma^2}\parallel \mathbf{y}-\mathbf{H}\mathbf{x} \parallel^2)}{\sum_{\mathbf{x}\in\mathcal{C}:b_{i}=0}\text{exp}\ (-\frac{1}{2\sigma^2}\parallel \mathbf{y}-\mathbf{H}\mathbf{x} \parallel^2)}.
\label{eqn:LLR derandomized LSD}
\end{equation}

It is noteworthy that lattice points with sampling probabilities $P(\mathbf{x})\geq\frac{1}{2K}$
will be deterministically sampled by derandomized sampling algorithm. As shown in
(\ref{distance2}), this can be interpreted as obtaining all the lattice points inside
a sphere of the radius $r$ where
\begin{equation}
r=\|\mathbf{y}-\mathbf{H}\mathbf{x}\|\leq \text{min}_ir_{i,i}\cdot\sqrt{\text{log}_\rho(2Ke^{-2n/\rho})}.
\label{distance7}
\end{equation}
To achieve a better upper bound of $r$, the optimum choice of the parameter $\rho_o$ in
(\ref{relationship K and rho}) is applied and we have
\begin{equation}
r\leq\sqrt{\frac{2n}{\rho_{\text{o}}}}\ \text{min}_ir_{i,i}.
\label{distance3}
\end{equation}

Let $\mathcal{C}_1$ denotes the set formed by lattice points within sphere radius $r$
\begin{equation}
\mathcal{C}_1\triangleq\{\mathbf{x}\in\mathcal{X}^n:\|\mathbf{y}-\mathbf{H}\mathbf{x}\|\leq r\},
\label{distance4}
\end{equation}
then set $\mathcal{C}$ in (\ref{eqn:LLR derandomized LSD}) can be rewritten as
\begin{equation}
\mathcal{C}=\mathcal{C}_1\cup\mathcal{C}_2,
\end{equation}
where $\mathcal{C}_2$ represents the set of lattice points outside radius
$r$ but also sampled by derandomized sampling decoding, and normally
$|\mathcal{C}_1|<|\mathcal{C}_2|$. Although lattice points within $r$ only
constitute a small part in the final candidate list of derandomized sampling, it captures
the key aspect of the decoding performance and offers a way to investigate
the effect of the sample size $K$ in soft-output decoding.
\begin{figure*}
\setcounter{figure}{4}
\begin{center}
\includegraphics[width=4.4in]{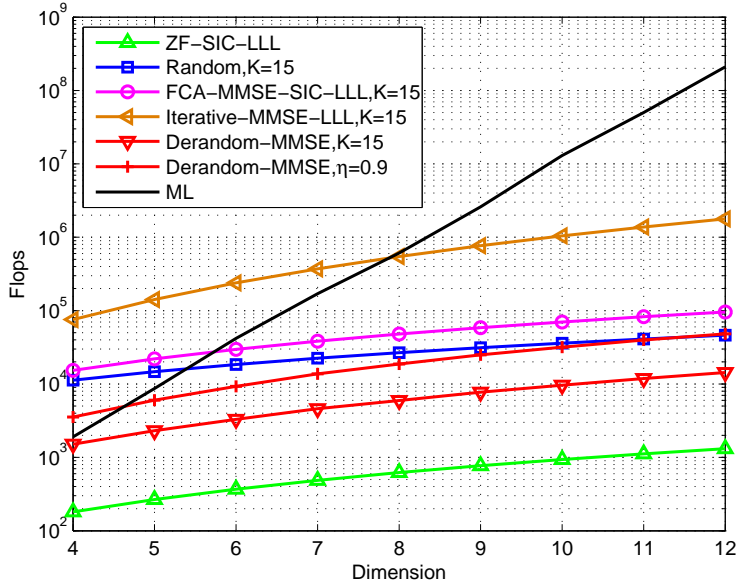}\\
\end{center}
\vspace{-1em}
\caption{Complexity comparison in flops for a MIMO system using 64-QAM at SNR per bit = 17dB.}
\label{simulation 4}
\end{figure*}

Based on the upper bound of the sphere radius $R$ in LSD, derandomized sampling algorithm
can also be applied to implement soft-out decoding through sampling all the lattice points within $R$.
Hence, according to (\ref{eqn:upper bound of LSD}), let $r=\sqrt{n}\sigma$, the derandomized
sampling algorithm will achieve near-MAP performance even with lattice points in
$\mathcal{C}_1$ only. Therefore, we have
\begin{equation}
\sqrt{\frac{2n}{\rho_{\text{o}}}}\ \text{min}_ir_{i,i}=\sqrt{n}\sigma
\label{distance5}
\end{equation}
and
\begin{equation}
\rho_{\text{o}}=\frac{2\ \text{min}_i^2r_{i,i}}{\sigma^2}.
\label{rho}
\end{equation}

By substituting (\ref{rho}) into (\ref{relationship K and rho}), the
corresponding sample size $K$ can be derived as
\begin{equation}
K=\frac{1}{2}\left(\frac{2e\hspace{0.1em}\text{min}_i^2r_{i,i}}{\sigma^2}\right)^{\left(\frac{n\sigma^2}{\text{min}_i^2r_{i,i}}\right)}.
\label{final K}
\end{equation}

Obviously, with the increment of $K$, more lattice points will be
sampled and the corresponding sphere radius $r$ also
increases gradually leading to further performance improvement. Note that
the total sampling probability shown in (\ref{new lower bound}) could also
be used to reveal this flexible trade-off. As for achieving near-MAP performance,
we emphasize that the required sample size $K$ of the derandomized sampling
algorithm is significantly less than the value
shown in (\ref{final K}). The reasons are two-fold: the derivation is based on a
loose upper bound of $r$ shown in (\ref{distance7}), and the
contribution of set $\mathcal{C}_2$ which also contains sampled lattice points is
not considered. Nevertheless, it provides a straightforward
way of showing the trade-off in soft-output decoding with respect to $K$. Actually,
with a moderate value of $K$, desirable performance gain in a low complexity burden
can be achieved, as will be shown in simulation results.

\section{Simulation Results}
In this section, performance and complexity of the derandomized
sampling algorithm in MIMO systems are studied. Here, the $i$-th
entry of the transmitted signal $\textbf{x}$, denoted as $x_i$, is the modulation symbol
taken independently from a $Q^2$-QAM constellation $\mathcal{X}$
with Gray mapping. We assume a flat fading environment, the channel matrix
$\mathbf{H}$ contains uncorrelated complex gaussian fading gains with unit
variance and remains constant over each frame duration. Let $E_b$ represents the
average power per bit at the receiver, then $E_b/N_0=n/(\text{log}_2(M)\sigma^2)$
holds where $M$ is the modulation level and $\sigma^2$ is the noise power.

Fig. \ref{simulation 1} shows the bit error rate (BER) of the derandomized sampling
decoding compared with other decoding schemes in a $10\times10$ uncoded MIMO system
with 64-QAM. Clearly, sampling decoding schemes have considerable
gains over the lattice-reduction-aided SIC. Compared to the fixed candidates
algorithm (FCA) in \cite{MaxiaoliSoft} and iterative list decoding in
 \cite{Shimokawa} with 30 samples, sampling decoding algorithms offer not only
the improved BER performance but also the promise of smaller sample
size. As expected, derandomized sampling decoding achieves better BER performance than randomized sampling decoding with the
same $K$. Specifically, the gain in MMSE schemes
with $K=15$ is approximately 1 dB for a BER of $10^{-4}$.
With the increment of $K$, the BER performance improves
gradually. Observe that with $\eta=0.9$ ($K$=73), the performance of
the derandomized sampling algorithm suffers negligible loss compared with
ML. Therefore, with a moderate $K$, derandomized sampling decoding
could achieve near-ML performance.
\begin{figure*}
\setcounter{figure}{5}
\begin{center}
\includegraphics[width=4.4in]{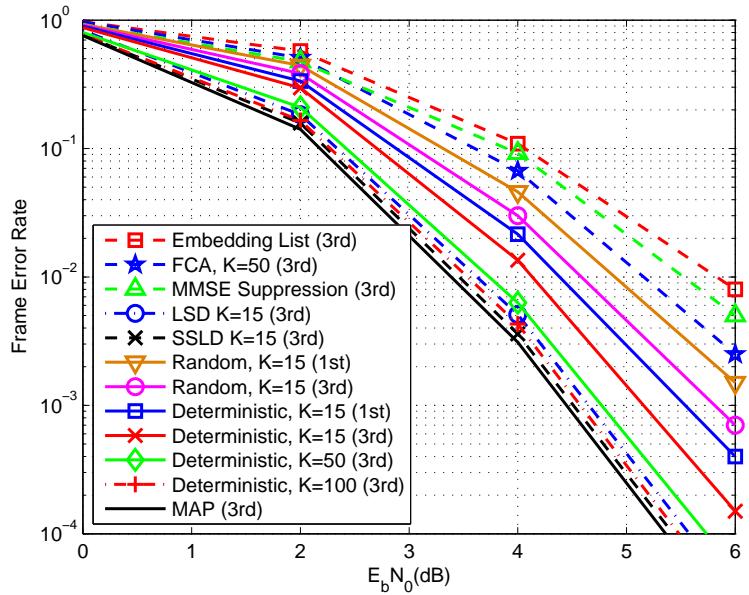}\\
\end{center}
\vspace{-1em}
\caption{Frame error rate versus average SNR per bit in the coded $8 \times 8$ MIMO BICM-IDD system with a rate-1/2 LDPC code of codeword length 256 using 4-QAM.}
\label{simulation 6}
\end{figure*}

Fig. \ref{simulation 4} shows the complexity comparison of the
derandomized sampling algorithm with other schemes in different
dimensions. It is clearly seen that in a 64-QAM MIMO system for the
fixed SNR ($E_b/N_0=17$ dB), the derandomized sampling algorithm
needs much lower average flops than other decoding schemes with the
same size $K$. This can be interpreted as reducing the computation
in sampling procedures by removing all the unnecessary repetitions.
Even for a large $K=73$ $(\eta=0.9)$, the complexity is still
lower than that of the randomized sampling algorithm with $K=15$. Consequently,
better BER performance and less complexity requirement make derandomized sampling
algorithm very promising for lattice decoding.

Fig. \ref{simulation 6} shows the frame error rate for a coded $8 \times 8$
MIMO BICM-IDD system with 4-QAM, using a rate-1/2, irregular (256,128,3) low-density
parity-check (LDPC) code of codeword length 256 (i.e., 128 information bits).
Each codeword spans one channel realization and a random bit interleaver is used.
The parity check matrix is randomly constructed, but cycles of length 4 are eliminated.
The maximum number of decoding iterations for LDPC is set at 50. Clearly,
after three iterations between MIMO detector and SISO decoder in IDD, the proposed
sampling algorithm with $K=15$ performs better than FCA, MMSE suppression \cite{Matache},
and Embedding list \cite{EmbeddingLuzzi}. To achieve a better comparison, performance of both sampling
algorithms with $K=15$ after one iteration are also given. As expected, derandomized sampling algorithm always achieves better FER performance than randomized sampling algorithm under the
same iteration. Note that there is no significant performance
gain after more than three iterations in IDD receivers. It is observed that the LSD
in \cite{Hochwald} and shifted sphere list deocding (SSLD) in \cite{BoutrosSoft} with sample size $K=15$ achieve near-MAP performance. However, due to the
application of sphere decoding, their complexity are high and increase exponentially
with the size of sphere radius and system dimensions leading them impractical.
It is also shown that the performance gap between the proposed algorithm and MAP
decreases with the increment of $K$ and near-MAP performance is achieved
by derandomized sampling algorithm with a moderate size $K=100$. By adjusting $K$,
the whole system enjoys a flexible trade-off between performance and complexity.

\section{Conclusions}
In this paper, we proposed a derandomized algorithm to address
issues in sampling algorithms caused by randomization, which holds
great potential in both lattice decoding and soft-output decoding.
By setting a probability threshold to perform the sampling, the whole
sampling procedure becomes deterministic. We demonstrated that
the proposed derandomized sampling algorithm outperforms the randomized sampling
algorithm with much lower complexity and derived the optimal parameter
$A$ which maximizes the decoding radius $R_{\text{Derandomized}}$ for the
best decoding performance. To accomplish the trade-off in lattice decoding,
the upper bound on the sample size $K$ corresponding to near-ML performance was also given.
Furthermore, we found that the proposed derandomized sampling algorithm is quite suitable
for soft-output decoding through sampling a list of lattice points around the
ML estimate. According to the analysis, we demonstrated that the derandomized
sampling algorithm is capable of achieving near-MAP performance with a moderate
size $K$. Therefore, by varying $K$, the decoder enjoys a flexible trade-off
between performance and complexity in both lattice
decoding and soft-output decoding.

\bibliographystyle{IEEEtran}
\bibliography{IEEEabrv,reference1}

\begin{thebibliography}{10}
\providecommand{\url}[1]{#1}
\csname url@samestyle\endcsname
\providecommand{\newblock}{\relax}
\providecommand{\bibinfo}[2]{#2}
\providecommand{\BIBentrySTDinterwordspacing}{\spaceskip=0pt\relax}
\providecommand{\BIBentryALTinterwordstretchfactor}{4}
\providecommand{\BIBentryALTinterwordspacing}{\spaceskip=\fontdimen2\font plus
\BIBentryALTinterwordstretchfactor\fontdimen3\font minus
  \fontdimen4\font\relax}
\providecommand{\BIBforeignlanguage}[2]{{%
\expandafter\ifx\csname l@#1\endcsname\relax
\typeout{** WARNING: IEEEtran.bst: No hyphenation pattern has been}%
\typeout{** loaded for the language `#1'. Using the pattern for}%
\typeout{** the default language instead.}%
\else
\language=\csname l@#1\endcsname
\fi
#2}}
\providecommand{\BIBdecl}{\relax}
\BIBdecl

\bibitem{Babai}
L.~Babai, ``On {L}ov\'asz' lattice reduction and the nearest lattice point
  problem,'' \emph{Combinatorica}, vol.~6, no.~1, pp. 1--13, 1986.

\bibitem{Luzzi}
L.~Luzzi, G.~Othman, and J.~Belfiore, ``Augmented lattice reduction for
  low-complexity {MIMO} decoding,'' \emph{IEEE Trans. Wireless Commun.},
  vol.~9, pp. 2853--2859, Sep. 2010.

\bibitem{WubbenMMSE}
D.~Wubben, R.~Bohnke, V.~Kuhn, and K.~D. Kammeyer, ``Near-maximum-likelihood
  detection of {MIMO} systems using {MMSE}-based lattice reduction,'' in
  \emph{Proc. IEEE Int. Conf. Commun.(ICC'04)}, Paris, France, Jun. 2004, pp.
  798--802.

\bibitem{Agrell2002}
E.~Agrell, T.~Eriksson, A.~Vardy, and K.~Zeger, ``{Closest} point search in
  lattices,'' \emph{IEEE Trans. Inform. Theory}, vol.~48, no.~8, pp.
  2201--2214, Aug. 2002.

\bibitem{Gan2009}
Y.~H. Gan, C.~Ling, and W.~H. Mow, ``{Complex} lattice reduction algorithm for
  {low-complexity full-diversity MIMO detection},'' \emph{IEEE Trans. Signal
  Process.}, vol.~57, no.~7, pp. 2701--2710, Jul. 2009.

\bibitem{JaldenDMTJ}
J.~Jalden and P.~Elia, ``{DMT} optimality of {LR}-aided linear decoders for a
  general class of channels, lattice designs, and system models,'' \emph{IEEE
  Trans. Inform. Theory}, vol.~56, no.~10, pp. 4765--4780, Oct. 2010.

\bibitem{CongProxity}
C.~Ling, ``On the proximity factors of lattice reduction-aided decoding,''
  \emph{IEEE Trans. Signal Process.}, vol.~59, no.~6, pp. 2795--2808, Jun.
  2011.

\bibitem{DamenDetectionSearch}
M.~O. Damen, H.~Gamal, and G.~Caire, ``On maximum-likelihood detection and the
  search for the closest lattice point,'' \emph{IEEE Trans. Inform. Theory},
  vol.~49, pp. 2389--2401, Oct. 2003.

\bibitem{Hochwald}
B.~M. Hochwald and S.~ten Brink, ``Achieving near-capacity on a
  multiple-antenna channel,'' \emph{IEEE Trans. Commun.}, vol.~51, no.~3, pp.
  389 -- 399, Mar. 2003.

\bibitem{Hassibisoft}
H.~Vikalo, B.~Hassibi, and T.~Kailath, ``Iterative decoding for {MIMO} channels
  via modified sphere decoding,'' \emph{IEEE Trans. Wireless Commun.}, vol.~3,
  no.~6, pp. 2299 -- 2311, Nov. 2004.

\bibitem{Silvolasoft}
P.~Silvola, K.~Hooli, and M.~Juntti, ``Suboptimal soft-output map detector with
  lattice reduction,'' \emph{IEEE Signal Process. Lett.}, vol.~13, no.~6, pp.
  321 -- 324, Jun. 2006.

\bibitem{MaxiaoliSoft}
W.~Zhang and X.~Ma, ``{Low-complexity soft-output decoding with
  lattice-reduction-aided detectors},'' \emph{IEEE Trans. Commun.}, vol.~58,
  no.~9, pp. 2621--2629, Sep. 2010.

\bibitem{Millinersoft}
D.~Milliner and J.~Barry, ``A lattice-reduction-aided soft detector for
  multiple-input multiple-output channels,'' in \emph{Proc. IEEE Globecom'06},
  Nov. 2006, pp. 1--5.

\bibitem{BoutrosSoft}
J.~Boutros, N.~Gresset, L.~Brunel, and M.~Fossorier, ``Soft-input soft-output
  lattice sphere decoder for linear channels,'' in \emph{Proc. IEEE GLOBECOM},
  San Francisco, Dec. 2003, pp. 1583--1587.

\bibitem{Leesoft}
J.~Lee, B.~Shim, and I.~Kang, ``Soft-input soft-output list sphere detection
  with a probabilistic radius tightening,'' \emph{IEEE Trans. Wireless
  Commun.}, vol.~11, no.~8, pp. 2848 --2857, Aug. 2012.

\bibitem{CongRandom}
S.~Liu, C.~Ling, and D.~Stehle, ``{Decoding by sampling: a randomized lattice
  algorithm for bounded distance decoding},'' \emph{IEEE Trans. Inform.
  Theory}, vol.~57, pp. 5933--5945, Sep. 2011.

\bibitem{Klein}
P.~Klein, ``Finding the closest lattice vector when it is unusually close,''
  \emph{SIAM Symposium on Discrete Algorithms}, pp. 937--941, ACM, 2000.

\bibitem{LarssonPartialMarginalization}
E.~Larsson and J.~Jalden, ``Fixed-complexity soft {MIMO} detection via partial
  marginalization,'' \emph{IEEE Trans. Signal Process.}, vol.~56, pp.
  3397--3407, Aug. 2008.

\bibitem{Renqiuwangsoft}
R.~Wang and G.~B. Giannakis, ``Approaching {MIMO} channel capacity with soft
  detection based on hard sphere decoding,'' \emph{IEEE Trans. Commun.},
  vol.~54, no.~4, pp. 587 -- 590, Apr. 2006.

\bibitem{Studersoft}
C.~Studer and H.~Bolcskei, ``Soft-input soft-output single tree-search sphere
  decoding,'' \emph{IEEE Trans. Inform. Theory}, vol.~56, no.~10, pp. 4827
  --4842, Oct. 2010.

\bibitem{HassibiPruning}
R.~Gowaikar and B.~Hassibi, ``Statistical pruning for near-maximum likelihood
  decoding,'' \emph{IEEE Trans. Signal Process.}, vol.~55, no.~6, pp.
  2661--2675, Jun. 2007.

\bibitem{ZhaoSD}
W.~Zhao and G.~Giannakis, ``Sphere decoding algorithms with improved radius
  search,'' \emph{IEEE Trans. Commun.}, vol.~53, no.~7, pp. 1104--1109, Jul.
  2005.

\bibitem{ShimSD}
B.~Shim and I.~Kang, ``Sphere decoding with a probabilistic tree pruning,''
  \emph{IEEE Trans. Signal Process.}, vol.~56, no.~10, pp. 4867--4878, Oct.
  2008.

\bibitem{Shimokawa}
T.~Shimokawa and T.~Fujino, ``Iterative lattice reduction aided {MMSE} list
  detection in {MIMO} system,'' in \emph{Proc. IEEE International Conference on
  Advanced Technologies for Communications}, Oct. 2008, pp. 50--54.

\bibitem{JaldenFSD}
J.~Jalden, L.~Barbero, B.~Ottersten, and J.~Thompson, ``{The error probability
  of the fixed-complexity sphere decoder},'' \emph{IEEE Trans. Signal
  Process.}, vol.~57, pp. 2711--2720, Jul. 2009.

\bibitem{Banaszczyk}
W.~Banaszczyk, ``New bounds in some transference theorems in the geometry of
  numbers,'' in \emph{Math. Ann. 296}, 1993, pp. 625--635.

\bibitem{Matache}
A.~Matache, C.~Jones, and R.~D. Wesel, ``Reduced complexity {MIMO} detectors
  for {LDPC} coded systems,'' in \emph{Proc. IEEE Military Commun. Conf.},
  Monterey, USA, 2004, pp. 1073--1079.

\bibitem{EmbeddingLuzzi}
L.~Luzzi, D.~Stehle, and C.~Ling, ``Decoding by embedding: correct decoding
  radius and {DMT} optimality,'' \emph{IEEE Trans. Inform. Theory}, vol.~59,
  no.~5, pp. 2960--2973, 2013.

\end{thebibliography}




\end{document}